\documentclass[a4paper]{jpconf}
\usepackage{graphicx}
\usepackage{subfigure}
\begin{document}
\title{The Heavy Photon Search experiment at Jefferson Laboratory}

\author{Andrea Celentano, for the HPS Collaboration}

\address{INFN-Genova, Via Dodecaneso 33, 16147 Genova, IT}

\ead{andrea.celentano@ge.infn.it}

\begin{abstract}
The Heavy Photon Search experiment (HPS) at Jefferson Laboratory will search for a new $U(1)$ massive gauge boson, or ``heavy-photon'', mediator of a new fundamental interaction, called ``dark-force'', that couples to ordinary photons trough kinetic mixing. HPS has sensitivity in the mass range 20 MeV - 1 GeV and coupling $\varepsilon^2$ between 10$^{-5}$ and 10$^{-10}$.  The HPS experiment will search for the $e^+e^-$ decay of the heavy photon, by resonance search and detached vertexing, in an electron beam fixed target experiment. 
HPS will use a compact forward spectrometer, which employs silicon microstrip detectors for vertexing and tracking, and a PbWO$_4$ electromagnetic calorimeter for energy measurement and fast triggering.

\end{abstract}

\section{Introduction}

The ``heavy-photon'', or $A^{\prime}$, is an hypothetical massive gauge boson associated to an additional $U(1)$ hidden symmetry, as predicted by several ``Beyond the Standard Model'' (BSM) theories \cite{Feldman1,Andreas1}. The $A^{\prime}$ couples to ordinary photons through a ``kinetic mixing'' mechanism \cite{Holdom,Galison}, i.e. through a loop-order effect, mediated by a GUT-scale massive particle that carries both the standard-model weak hypercharge and the dark-force equivalent.
This induces an effective parity-conserving interaction $\varepsilon e A^{\prime}_{\mu} J^{\mu}_{\mbox{\small EM}}$ of the $A^{\prime}$ to the electromagnetic current $J^{\mu}_{\mbox{\small EM}}$, with coupling $\varepsilon e$, where the parameter $\varepsilon$ can naturally be in the range 10$^{-12}$ - 10$^{-2}$ \cite{Goodsell1,Cicoli1,Essig1}. 
Consequently, standard model charged particles acquire a ``milli-charge'' $\varepsilon e$ for the interaction with the ``heavy-photon'', resulting in em-like processes such as $A^{\prime}$-strahlung (the emission of an $A^{\prime}$ by an high-energy electron) and the $A^{\prime}$ decay to electron-positron pairs.  

Recently, a possible role of heavy photons in dark matter physics has been suggested, generating great interest
to directly search for the $A^{\prime}$ \cite{Arkani,Pospelov1}. Results from indirect and direct dark matter searches have been re-interpreted as potential signals of dark matter interacting with heavy photons.
For example, the positron excess in the cosmic-ray flux recently reported by AMS \cite{AMS}, confirming previous results from PAMELA \cite{PAMELA1} and FERMI \cite{FERMI}, can be explained assuming that dark matter is charged under the ``dark-force'' interaction, and decays or annihilates to $A^{\prime}$ particles, that in turn decay to $e^{+} e^{-}$ pairs. In this picture, the $A^{\prime}$ acts as a ``mediator'' between the Standard Model and the dark matter sector. The non-observed excess in anti-proton fraction suggests that the $A^{\prime}$ mass is lower than $\simeq$ 2 GeV \cite{PAMELA2}.

In the simplest scenarios, the two main parameters that determine the characteristics of the $A^{\prime}$ and thus the experimental
search strategies are: the kinetic mixing parameter $\varepsilon$ and the mass $M_A$. 
While a huge range of mixing parameters and masses are possible, it is natural that $\varepsilon$ be around $10^{-3}$, and necessary that masses be around GeV, if the positron excess is to be explained. Other than explaining astrophysics anomalies, an $A^{\prime}$ in this mass and coupling range can explain the muon anomalous magnetic moment \cite{Pospelov2}.
This parameter space region has been investigated by a large number of experiments, with so far no positive results (see \cite{Snowmass} for a complete review). However, a significant part of it is still unknown and will be explored by planned and proposed experiments. 

High luminosity, fixed target experiments provide an unique way to probe this unknown region \cite{Bjorken1}. In these experiments, an electron beam impinges on a high Z target, producing heavy photons that subsequently decay to pairs of fermions. The $A^{\prime}$ is then reconstructed by measuring the decay products momenta.

\begin{figure}[tpb]
\centering
\subfigure{\includegraphics[width=0.48\textwidth]{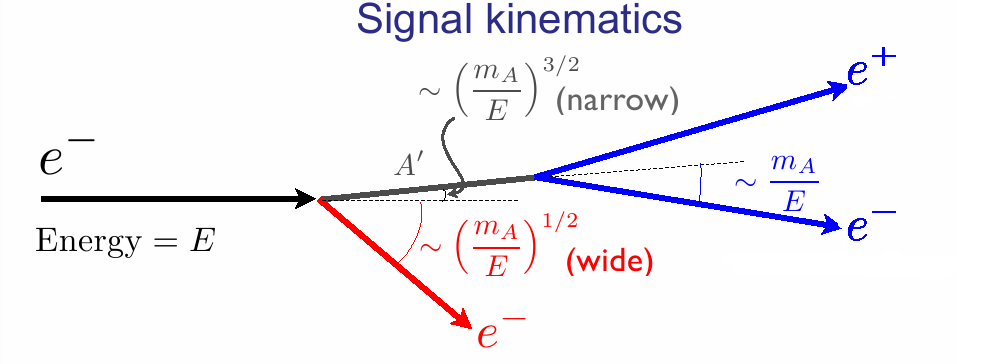}}
\subfigure{\includegraphics[width=0.48\textwidth]{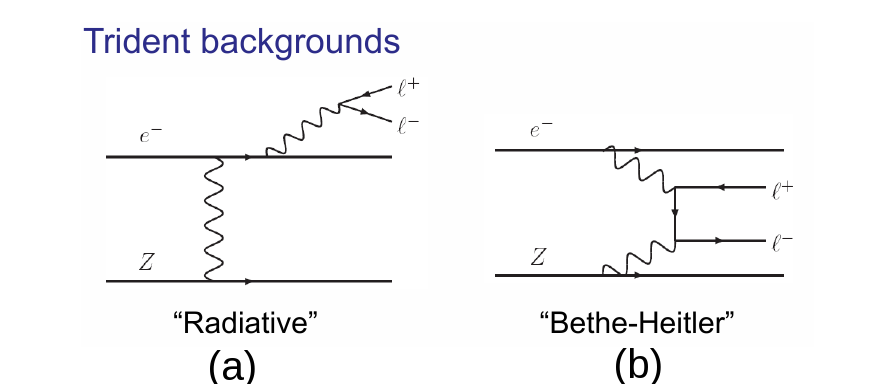}}
\caption{\small Left: the foreseen kinematics for the $A^{\prime}$ production by Bremsstrahlung off an incoming electron and the subsequent decay to an $e^{+} e^{-}$ pair. Right: sample diagrams of radiative trident ($\gamma^{*}$) and Bethe-Heitler trident reactions that comprise the primary background to the $A^{\prime}$ search. \label{fig:kin}}
\end{figure}

The kinematic of this process, sketched in Figure \ref{fig:kin} (left panel), is characterized by a very-forward emitted $A^{\prime}$ that carries most of the incident electron beam energy $E$. This results in a very small opening angle of the decay products, of the order of $\theta \simeq M_{A}/E$. 
Therefore, the $A^{\prime}$ decay products will be emitted at a very low angle with respect to the beam-line, and will carry most of the beam energy. 
Furthermore, depending on the value of the parameters, long-lived $A^{\prime}$ are possible in the aforementioned parameter space, with displaced vertexes up to hundreds of meters.

The main background sources associated to this reaction are QED trident processes, i.e. Bethe-Heitler trident reactions and radiative trident production (see Figure \ref{fig:kin}, right-panel). Although the Bethe-Heitler process has a much larger total cross-section, it can be significantly reduced by exploiting its very different kinematics. 
Instead, the radiative trident process is an irreducible background, with the same kinematic as the signal. However, it presents a smooth, continuous distribution in the $e^{+} e^{-}$ invariant mass distribution, in contrast to the peak at $M_{e^+e^-}=M_A$ for the signal, and occurs promptly at the production target, while the signal can be displaced for small $\varepsilon$ values.

The Heavy Photon Search (HPS) experiment at Jefferson Laboratory \cite{HPS} is a fixed-target experiment specifically designed to search for dark-photons in the mass range between 20 MeV and 1 GeV and coupling $\varepsilon^{2}$ between $10^{-5}$ and $10^{-10}$. HPS will search for the $e^{+}e^{-}$ decay of the heavy photon, through complementary techniques: resonance search (the traditional ``bump-hunting'') and detached vertexing. In particular, HPS will explore a unique region corresponding to small cross sections, out of reach from collider experiments, where thick absorbers, as used in beam-dump experiments, are not allowed due to the relatively short $A^{\prime}$ decay length ($<$ 1 m).

\section{The HPS experiment at Jefferson Laboratory}

The HPS experiment will run in Hall B of Thomas Jefferson National Accelerator Facility (JLab). The high duty-cycle, high-quality electron beam, with energy tunable between 1.1 and 11 GeV and current up to 250 nA, will permit to reach high luminosity while keeping beam-related backgrounds under control.  
\begin{figure}[tpb]
\centering
\includegraphics[width=0.85\textwidth]{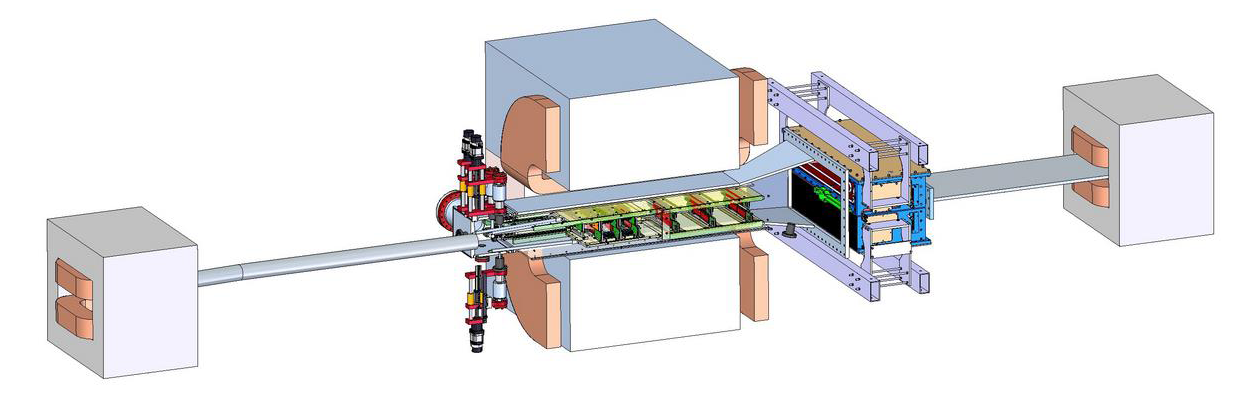}
\caption{\small \label{fig:setup} Schematic of the HPS detector setup. From left to right: the first dipole magnet, the target system, the silicon vertex tracker (SVT) inside the analyzing dipole magnet, the electromagnetic calorimeter (ECal), the third dipole magnet.}
\end{figure}
The HPS detector was designed to match the foreseen signal kinematics described above. Mandatory requirements are: a large and uniform acceptance in the forward region close to the beam, down to $\theta \simeq 15$ mrad, excellent track reconstruction efficiency for electrons and positrons, good angular and momentum resolution to reconstruct invariant mass precisely, and excellent vertex resolution to discriminate displaced $A^{\prime}$ decays from prompt QED backgrounds. Radiation hardness is also a critical issue, given the high radiation environment, particularly near the target position. Finally, the material budget must be as low as possible, to minimize multiple scattering and preserve the angular resolution.

The HPS detector is made of two main sub-components, a $\simeq$ 1 m long silicon tracker inside a dipole magnet (SVT), to measure charged particle trajectories and vertexes, and a fast lead-tungstate electromagnetic calorimeter (ECal), to measure particle energies, identify electrons and provide a fast trigger signal. The detector setup is shown in Figure \ref{fig:setup}. HPS employs a three-magnet system with the second dipole serving as an analyzing magnet. The tungsten target is located in front of the analyzing magnet, at about 10 cm from the first SVT layer.
To reduce beam-related backgrounds, the target and the SVT are mounted in a vacuum chamber. The ECal is mounted outside the vacuum chamber.

The SVT is made of 6 layers of Si modules, each with two sensors. The introduction of a small stereo angle in between the two (50 or 100 mrad) provides three-dimensional tracking and vertexing.  Each layer in one half is supported on a common support plate with independent cooling and readout. The SVT is split in two modules, mounted on top and on bottom of the beam plane, leaving a central ``dead-region'' where the degraded beam can pass through the detector unobstructed. 
Sensors are readout by the APV25 chip, developed for the CMS experiment at the CERN LHC \cite{APV}, with a custom DAQ system based on the Advanced Telecom Communications Architecture (ATCA) technology.

The electromagnetic calorimeter is also split into two halves. Each half of the ECal consists of 221 PbWO$_4$ crystals arranged in five rows. Each row is made of 46 crystals, expect the one closest to the beam that has only 37. 
The light from each crystal is read out by a Large Area Avalanche Photodiode (APD) glued on the back surface. Signals from the APDs are read using custom-made ampliﬁer boards before being sent to the data acquisition electronics. Each signal is acquired by a 250 MHz, 12 bit Flash Analog to Digital Converter (FADC), with 8 $\mu$s latency. The on-board FPGA processes the digitized samples to provide information about the energy and the timing associated with each hit.

\subsection{HPS reach}

\begin{figure}[tpb]
\centering
\includegraphics[width=0.45\textwidth]{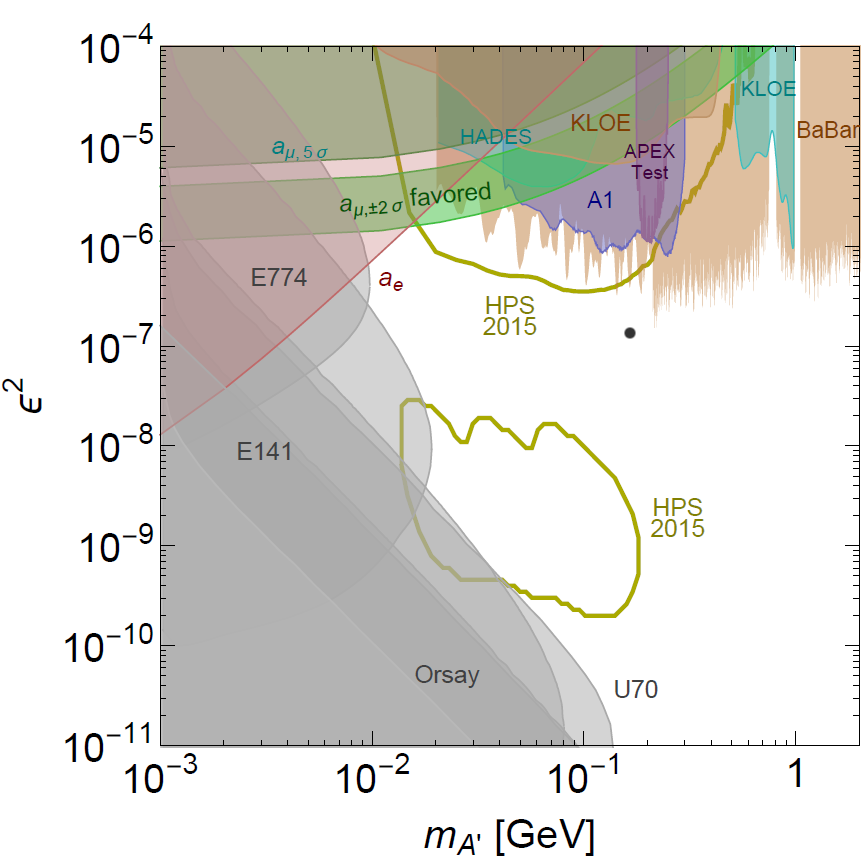}
\caption{\small The HPS projected reach. The continuous yellow line corresponds to the 2$\sigma$ limit. The preferred muon g-2 region is reported, together with published exclusion limits from other experiments. The newest results reported by the PHENIX collaboration are not included in this plot \cite{PHENIX}. \label{fig:reach}}
\end{figure}

The projected HPS reach is reported in Figure \ref{fig:reach}, together with published exclusion limits from other experiments. The green band $a_{\mu,\pm 2\sigma}$ ($a_{\mu,\pm 5\sigma}$) corresponds to the region where the $A^{\prime}$ can explain the discrepancy between the experimental value of the muon anomalous magnetic moment, considering a 2$\sigma$ (5$\sigma$) error, and the corresponding Standard Model calculation. The red band $a_e$, instead, corresponds to the exclusion region from the electron magnetic momentum.

The reach was computed by assuming a one-week run at 1.1 GeV, a one-week run at 2.2 GeV, and a two-weeks run at 4.4 GeV. The two regions in the reach plot correspond to the two analysis techniques employed in the experiment: the resonance search (at larger $\varepsilon^2$) and the detached vertex search (at lower $\varepsilon^2$). They are not connected since a minimum displacement of $\simeq 20$ mm at 2.2 GeV ($\simeq$ 10 mm at 6.6 GeV) is required for the vertex-based search.

\end{document}